# A Graph Computation based Sequential Power Flow Calculation for Large-Scale AC/DC Systems

Wei Feng, Jingjin Wu, Chen Yuan, Guangyi Liu, Renchang Dai
GEIRI North America Inc
Santa Clara, USA

Qingxin Shi, Fangxing Li
Department of Electrical Engineering & Computer Science
The University of Tennessee
Knoxville, USA

*Abstract*—This paper proposes a graph computation based sequential power flow calculation method for Line Commutated Converter (LCC) based large-scale AC/DC systems to achieve a high computing performance. Based on the graph theory, the complex AC/DC system is first converted to a graph model and stored in a graph database. Then, the hybrid system is divided into several isolated areas with graph partition algorithm by decoupling AC and DC networks. Thus, the power flow analysis can be executed in parallel for each independent area with the new selected slack buses. Furthermore, for each area, the node-based parallel computing (NPC) and hierarchical parallel computing (HPC) used in graph computation are employed to speed up fast decoupled power flow (FDPF). Comprehensive case studies on the IEEE 300-bus, polished South Carolina 12,000-bus system and a China 11,119-bus system are performed to demonstrate the accuracy and efficiency of the proposed method.

## I. Introduction

The AC/DC system, also called hybrid system are a promising solution for large capacity and long-distance transmission network with low power losses. To meet the requirements of modern power systems, the line commutated converter (LCC) based DC technology is employed to connect multiple distant AC grids and construct the large-scale hybrid systems. LCC, comparing to voltage source converter (VSC) is advantageous with higher transmission voltage and larger transmission capacity. LCC based HVDC projects have been built in the past decades [1] and are expected to be built more in the future [2]. Power flow analysis on the AC/DC hybrid system is fundamental for system planning, operation, and control. In general, there are two wide-used methods to solve the power flow of AC/DC systems, subdivided in unified [3] and sequential methods [4]. In the unified method, the variables and equations of AC and DC are integrated and calculated together. In the sequential method, correspondingly, the AC and DC equations are solved separately in a sequential process. One of the sequential approach advantages is the flexibility of implementing DC solver in addition to the existing AC power flow algorithms and software. In the unified method, however, the whole AC power flow calculation software needs to be reprogrammed to accommodate DC network solution.

Although studies have been done on the sequential method focusing on modeling DC grids, adding converter loss [5], and improving convergence [6], while few methods are focused on the computing performance. In the past decades, the scale of AC/DC system has been increasing dramatically, with more DC lines putting into use. Computation efficiency is becoming a challenge when the sequential method is applied to solve AC/DC system because of its lower convergence and the sequential iterations [7]. One of the promising methods to improve performance is parallel computing utilizing multi-threads of CPU or GPU with parallel computing capability [8]. However, the state-of-art of parallel power flow algorithms do not fully utilize the parallel capability because of the limitation of algorithms based on relational database. With the development of graph computation, a protogenetic parallel database and algorithm structure is constructed to speed up the power system analysis. In power system modeling, the complex system is stored intuitively as a 'graph' with edges and vertices to support NPC and HPC easily. Previous works have demonstrated the efficient parallel performance of using graph database and graph computing in power system applications, such as state estimation, power flow analysis, and contingency analysis [9]-[11]. The advantages of parallel computing can be further taken by dividing the AC/DC systems to separated sub-systems.

In this paper, a graph computation based sequential method is proposed to divide the system into blocks and speed up the power flow analysis in each block in parallel for large-scale AC/DC system without compromising accuracy. Firstly, a hybrid system is divided into several sub-systems with graph partition algorithms. The sub-systems are independent and decoupled by DC connections. Then, the power flow analysis of sequential method can be conducted in parallel for each sub-system. Furthermore, graph computation based FDPF is applied to solve power flow in parallel in the sub-system level in each iteration to achieve computation time-saving at the greatest extent.

The remainder of this paper is organized as follows. Section II gives a brief description of graph computing and its application in power system. Section III discusses the graph computation based sequential method and computation performance improvement. Section IV presents case studies on large-scale systems to verify the efficiency of the proposed method. Section V provides the conclusions.

## II. Graph Computing and Its Application

### A. Graph Database and Graph Modeling

In graph database, a power system is modeled as a graph G (V, E), where $v_i \in V$ represents the $i^{th}$ element of the vertex set

This work is supported by the State Grid Corporation technology project 5455HJ180020.

$V$, and $e_{ij} \in E$ denotes the link between vertex $i$ and $j$ of the edge set $E$. Parameters describing vertices and edges are stored as $P_{vi}, P_{eij} \in P$. Using graph data structure, the traditional relational data is expressed by graph modeling properties of vertices and edges and their topological connectivity.

In power systems, parameters of generators, loads, SVCs, STATCOMs, and other one-terminal components are defined as vertex attributes. Transmission lines, transformers, filters, breakers, and other multi-terminal components connecting in adjacent vertices are defined as edges. The system topology is then self-defined by the vertices and edges connectivity.

*B. Graph Computing*

Graph database and graph model support node-based parallel computing and hierarchical parallel computing.

*1) Node-based Parallel Computing.* NPC means each independent node can execute local computation at the same time. Forming Y-matrix is an example of node based parallel computing as shown in Fig. 1. In matrix *A*, the diagonal elements represent all nodes in the graph, the non-zero off-diagonal element indicates there is a connection between corresponding nodes, and the zero off-diagonal elements denote that there is no connection between two nodes. The non-zero off-diagonal element is calculated with attributes of the corresponding edge, and the diagonal element is calculated with attributes from the corresponding vertex and all connected edges. The calculations for diagonal elements are independent from one node to another. Therefore, they can be computed in nodal parallel.

*2) Hierarchical Parallel Computing (HPC).* HPC conducts calculation for nodes at the same level in parallel. After all the nodes in the lower level are calculated, the calculation for the nodes at the upper level will start in parallel after. One of the wide-used applications of HPC is LU factorization, which takes up a lot of time in solving power flow equations.

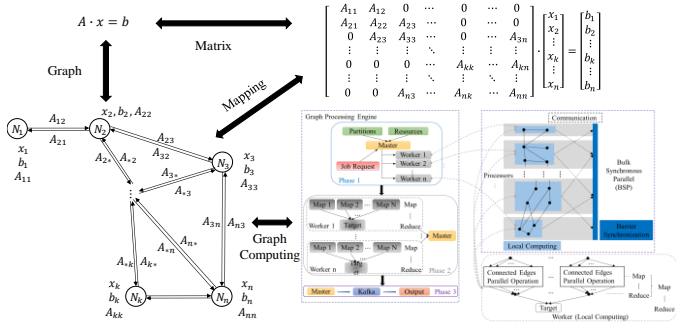

Figure 1. Application of Graph Computing in creating admittance matrix

*C. Graph Partition*

The graph partition is an optimal process to divide the original graph, $G_0(V, E)$ into smaller sub-graphs $G_1(V_1, E_1)$, $G_2(V_2, E_2), \ldots, G_k(V_k, E_k)$ in parallel to minimize the total cut edges between sub-sets, subject to the sizes of sub-graphs are closed to equal as shown as follow.

$$\begin{cases} |V_i| \approx \dfrac{|V|}{k}, \ i=1,...,k \\ \min_{P=(V_1,...,V_k)} |\partial_e P| \end{cases} \quad (1)$$

where *e* stands for cut edges. Cut edges, which also known as bridges in graph theory, together with the sizes are important in graph partitioning. Despite that a graph with *n* vertices may contain at least *(n-1)* cut edges, not all these edges can be selected as cut edges. The most optimal result is that each isolated area is connected via only one edge. Consider the complexity of real graph, several edges may exist between areas. Thus, the partition is transformed into an optimization problem which minimizes the total number of cut edges. Meanwhile, to fully take advantages of parallel computing threads, the number of isolated areas should coordinate the threads. Moreover, to avoid the waiting time between the largest and smallest area, the partitioned sizes should be close. Fig. 2 illustrates a conceptual 4 sub-graphs partitions, and the left partition is superior to the right one. Taking advantage of the graph structure of power system model, the mature partition algorithms are used to generate sub-systems.

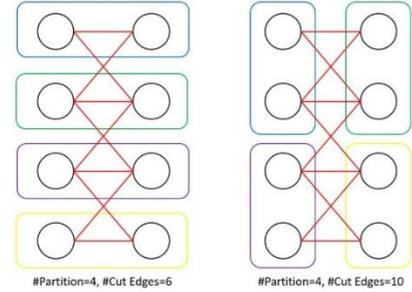

Figure 2. A Sample of graph partitioning

Generic graph partition approach is time-consuming because of the wide search range. In practice, taking account of power system architecture, DC lines and a few AC tie lines are usually used to connect separated AC systems. Cutting DC lines and tie lines are applied in this paper to partition AC/DC network.

### III. GRAPH COMPUTATION BASED POWER FLOW

*A. DC Model and Equations*

A typical LCC-based DC grid is composed of a rectifier station (R), an inverter station (I), converter transformers, filters, DC transmission lines, and other auxiliary devices. Since the thyristors in converter station are controlled by both triggering angle and the reverse voltage, R and I can be treated as mirror-image relation. The common power exchange model of AC/DC system is shown in Fig. 3.

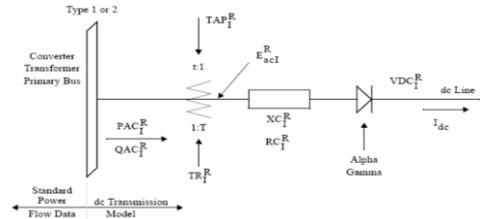

Figure 3. Power exchange model of AC/DC system (R/I)

To model the hybrid system, four sets of equations are built: AC grid equations, DC grid equations, AC/DC coupled (converter station) equations and control equations.

*1) AC grid equations.* As shown in Fig. 3, the left primary system is AC system, and the connected AC bus has injected power from DC side. The power mismatch equations are:

$$\Delta P_i = P_{i,inj} - V_i \sum_{j \in ii} V_j (G_{ij} \cos\theta_{ij} + B_{ij} \sin\theta_{ij}) - P_{DCR} = 0 \quad (2)$$

$$\Delta Q_i = Q_{i,inj} - V_i \sum_{j \in ii} V_j (G_{ij} \sin\theta_{ij} - B_{ij} \cos\theta_{ij}) - Q_{DCR} = 0 \quad (3)$$

where $P_{DCR}, Q_{DCR}$ stand for the injected power of R if $i^{th}$ bus is connected to DC grid. Since the structure of equations is the same with AC system, FDPF (fast decoupled power flow) can be used.

*2) DC grid equations.* For LCC grid in high-voltage system, the voltage-current equation is

$$V_{DC}^R = V_{DC}^I + I_{DC}R_{DC} \tag{4}$$

where $V_{DC}^{R,I}$ is for DC voltage of R/I, $I_{DC}$ stands for DC current and $R_{DC}$ is for DC line resistance.

*3) AC/DC coupled equations.* Take R side for instance, the injected active power and reactive power of R is:

$$P_{DCR} = V_{DC}^R I_{DC} \tag{5}$$

$$Q_{DCR} = P_{DCR} \tan\phi_R = V_{DC}^R I_{DC} \tan\phi_R \tag{6}$$

where $\phi_R$ is DC power factor of R. The parameters of a converter can be expressed as:

$$V_{DC}^R = N_r(\frac{3\sqrt{2}}{\pi}E_{AC}^R \cos\alpha - \frac{3}{\pi}X_{CR}I_{DC}) \tag{7}$$

$$\cos\phi_R = \cos\alpha - \frac{X_{CR}I_{DC}}{\sqrt{2}E_{AC}^R} \tag{8}$$

$$E_{AC}^R = V_i / T_R \tag{9}$$

where $N_r$ is the number of bridges, α for firing angle, $T_R$ for transformer turn ratio.

*4) Control equations.* To control the operating status of DC converters, both R and I are designed to work in two of the given operating statuses: constant DC power, constant DC voltage, constant DC current, constant angles (triggering or extinction), and constant transformer ratios. Since the number of equations equals the number of variables, all the parameters can be solved with simple calculation.

### B. Graph computation based sequential method

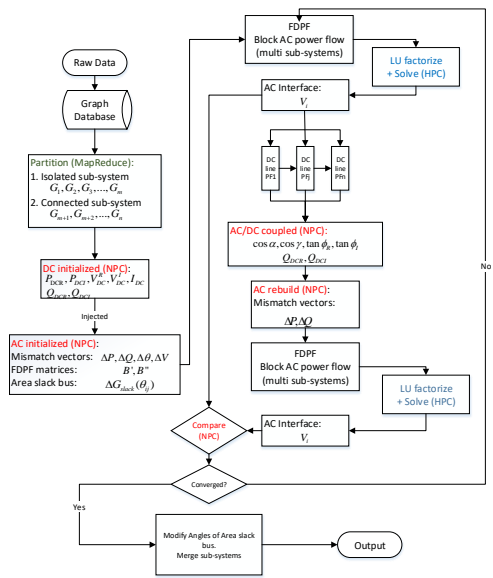

Figure 4. Flow chart of graph computation based sequential method

The sequential method calculates AC and DC sequentially and updates injected power for the next calculation until both systems achieve convergence. As shown in Fig. 4, graph partition, NPC and HPC can be utilized in the sequential method without changing the main calculation process. Firstly, the original hybrid system is partitioned to several sub-systems which could be solved independently with NPC and HPC. As discussed in previous sections, the calculation of building mismatch vectors $\Delta P/\Delta Q/\Delta\theta/\Delta V$, constructing the approximated Jacobian matrices $B'/B''$, and rebuilding vectors after one iteration can speed up with NPC. Furthermore, in each sub-system, the performance of solving FDPF for each iteration is improved with HPC.

To make better use of graph partition, the characteristics of hybrid system should be taken into consideration. In general, high-voltage LCC grids are used to connect several separate AC systems via ultra-long distances. Thus, AC/DC coupled buses (converter stations) and DC lines can be treated as the boundaries of sub-systems. In real hybrid systems, some low-voltage DC grid is built within a large local AC grid to enhance the stability when there exists high penetration of renewable energy. For such AC grid, the system network is still connected even the DC line is removed. Therefore, the goal of graph partition is to check if the sub-system is isolated when DC lines are cut.

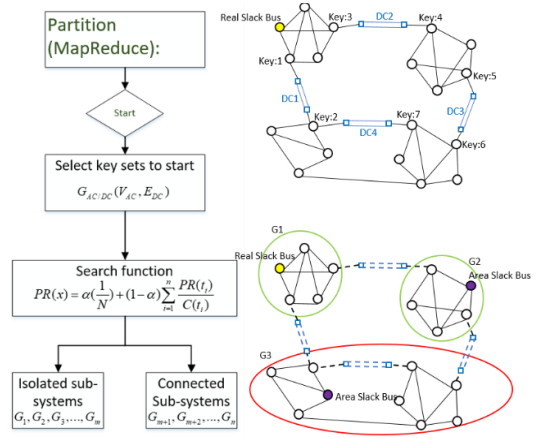

Figure 5. Flow chart of graph partition for hybrid system

For the purpose of illustration, assuming that the hybrid system is composed of three sub-systems, as shown in Fig. 5. After the raw data is converted to graph data, graph partition algorithms can be executed. The first step is to narrow the partition set to $G_{AC/DC}(V_{AC}, E_{DC})$, which only includes the coupled buses and DC lines. Then, each coupled bus conduct graph partition independently to generate sub-systems with the fewest cut edges except the DC lines. The sub-systems could be like G1/G2, which is fully isolated, or like G3 which is still connected via AC lines. Finally, select new area slack buses in each sub-system to do FDPF. The graph computation based FDPF can be implemented with NPC, which is used in initializing parameters, mismatch vectors and $B', B''$ matrices, and with HPC, which is used in factorizing LU and solving equations. The detailed applications have been discussed in the previous sections.

## IV. CASE STUDY AND RESULTS

### A. Test cases

In this section, three systems with different sizes and topologies are tested: the IEEE 300-bus system, an artificial 12,000-bus system, which is developed based on South Carolina

500-bus system, and a China 11,119-bus system. The standard IEEE 300-bus system is used to verify the accuracy of the proposed method. The parameters of the DC line connecting bus 119 and bus 120 are listed in Table I for verification test.

To demonstrate the computing performance, larger systems, 12,000-bus system and China 11,119-bus system, are tested. The 12,000-bus system is created based on South Carolina 500-Bus system [12]. It has 24 LCC lines (same as Table I) to connect the 24 sub-systems. The 11,119-bus system contains 9 LCC lines. Detailed parameters of its 9 LCC lines are shown in Table II. The other parameters are the same, with P-V control, reactance 7.936Ω, resistance 6.2Ω and transformer ratio 0.748.

TABLE I. PARAMETERS OF LCC LINE IN IEEE 300-BUS SYSTEM

| From Bus | To Bus | Bridges | Control | DC Power (MW) | DC Voltage (kV) |
|---|---|---|---|---|---|
| 119 | 120 | 4 | P-V | 100 | 460 |
| X(Ω) | R(Ω) | T ratio | α(degree) | γ(degree) | Block |
| 6.8 | 6.2 | 0.7478 | [15,20] | [18 20] | No |

TABLE II. PARAMETERS OF LCC IN 11,119-BUS SYSTEM

| From Bus | To Bus | DC Power(MW) | DC Voltage(kV) | α (degree) | γ (degree) |
|---|---|---|---|---|---|
| 88 | 1003 | 750 | 250 | [16 20] | [18 20] |
| 2957 | 4901 | 1500 | 500 | [15 20] | [18 20] |
| 3003 | 3717 | 1500 | 500 | [15 20] | [18 20] |
| 3520 | 4903 | 600 | 500 | [15 20] | [16 20] |
| 3528 | 4900 | 1500 | 500 | [16 20] | [16 20] |
| 7453 | 3746 | 360 | 225 | [18 20] | [18 20] |
| 7455 | 9491 | 750 | 250 | [12 20] | [12 20] |
| 7456 | 7772 | 1500 | 500 | [12 20] | [15 20] |

The testing environment is set up on TigerGraph, an efficient graph database which supports C, C++ and JAVA programming. The detailed configuration is shown in Table III.

TABLE III. GRAPH COMPUTATION ENVIRONMENT

| Software Environment | |
|---|---|
| Operating system | Red Hat 4.8.5 |
| Graph database | TigerGraph 2.1 |
| Hardware Environment | |
| CPU | 4×E7-8867 v3/2.50GHz |
| Memory | 64GB |

### B. Accuracy verification and computing performance

The power flow results of IEEE 300-bus system from PSS/E are taken as a benchmark to verify the accuracy of the proposed method, as shown in Table IV.

TABLE IV. ACCURACY VERIFICATION OF IEEE 300-BUS SYSTEM

| Test Platform | Bus | Magnitude (p.u.) | Phase (degree) | α/γ (degree) |
|---|---|---|---|---|
| TigerGraph (V2.1) | 119 | 1.0435 | 40.98738 | 16.240 |
| | 120 | 0.99818 | 37.72657 | 18.375 |
| PSS/E (V34.0) | 119 | 1.0435 | 40.9874 | 16.240 |
| | 120 | 0.99819 | 37.7266 | 18.379 |

TABLE V. TIME OF 12,000-BUS AND 11,119-BUS SYSTEMS (IN MS)

| Case | Iterations | 1 thread | 4 threads | 8 threads | 16 threads |
|---|---|---|---|---|---|
| 12,000 | 2 | 412.28 | 212.31 | 143.25 | 139.68 |
| 11,119 | 4 | 946.37 | 392.25 | 386.45 | 375.42 |
| Case | 24 threads | 32 threads | 64 threads | 96 threads | 128 threads |
| 12,000 | 136.53 | 137.12 | 151.25 | 157.68 | 156.43 |
| 11,119 | 362.21 | 352.49 | 372.34 | 376.53 | 395.43 |

The computation performance testing is conducted using the latter two systems with multi-threads, and the computing time is shown in Table V. To illustrate the improvement, the comparisons with the same sequential method developed based on Matpower in Matlab are given in Table VI. Both results are based on the best performance with the average of multiple tests, together with detailed time of two important processes. Since there exist multiple iterations in the sequential method, the time in Table VI is for the first iteration of creating matrices and factorizing LU.

TABLE VI. COMPARISONS OF 11,119-BUS AND 12,000-BUS SYSTEMS (IN MS)

| Platform | Case | Total | $B'$, $B''$ | LU |
|---|---|---|---|---|
| TigerGraph | 11,119 | 352.49 | 17.894 | 77.35 |
| | 12,000 | 136.53 | 23.84 | 65.25 |
| Matlab | 11,119 | 1491.5 | 61.3 | 367.2 |
| | 12,000 | 538.2 | 66.7 | 293.9 |

### C. Results analysis

As shown in Table IV, the largest mismatch of PF results is smaller than $1.0 \times 10^{-5}$ and the control angles are the same considering the discrete transformer tap. Thus, power flow accuracy is verified. Furthermore, the results of Table V demonstrate the promising capability of parallel computing with multiple threads. The best performances are achieved with the usages of 24 or 32 threads. In Table VI, compared with the convent sequential method, the proposed approach takes only 25.37% and 23.63% of computation time for 11,119-bus system and 12,000-bus system, respectively. Since the 11,119-bus system costs more iterations than the 12,000-bus system to converge, its computation time is longer. To reduce the influences of other essential calculations, the two main time-consuming segments, building matrices and factorizing LU, are also compared and shown in Fig. 6.

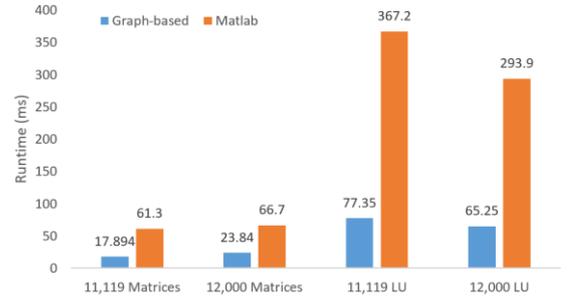

Figure 6. Speed-up gains of graph-based method

As shown in Fig. 6, the NPC based matrices building and HPC based LU factorization gain great speed-up compared with conventional methods. Taking advantages of multiple threads and the independent computing vertices, the computing performance can improve greatly. For a hybrid system with the size of over 10,000 buses, the average time saving with NPC and HPC can be 32.6% and 21.7% respectively.

TABLE VII. TIME CONSUMPTION OF 12,000-BUS SYSTEM (IN MS)

| Case | 11,119-Bus | | 12,000-Bus | |
|---|---|---|---|---|
| Threads | 1 | 24 | 1 | 24 |
| Graph Partition | 37.32 | 9.361 | 24.369 | 7.523 |
| DC Initialize | 9.231 | 6.723 | 7.65 | 5.652 |
| AC Initialize | 91.329 | 14.848 | 72.67 | 18.141 |
| LU Factorize (Total) | 656.496 | 205.241 | 131.787 | 75.6 |
| AC/DC Coupling | 12.699 | 14.372 | 3.760 | 3.922 |
| AC Rebuild | 38.233 | 5.321 | 36.52 | 6.21 |

Furthermore, to illustrate the utilization of NPC and HPC with multiple threads, the detailed computing time for two test systems, with 1 thread and 24 threads, are listed in Table VII. The calculating time of graph partition, NPC for AC/DC initialization and matrix rebuilding, HPC for LU factorization in solving power flow is listed. The usage of 24 threads presents a significantly better performance and the average time spending is 27.37%, 26.61 % and 35.63% for graph partition, NPC and HPC, compared with the computation using a single thread. The results denote that graph computation has a good capability of utilizing multiple threads to achieve better computing performance. However, it's also observed that further increase of threads put into calculation can only gain little speed-up, or even worse, leads to time increases. The abnormal results are due to the communication time between threads and RAM, and the overhead time cost in calling multiple threads. When the extra cost outweighs the time saved by the proposed graph computation, the total computation time will no longer reduce. Fig. 7 shows the computing time with different threads in 12,000-bus system, and the time comparison of the first inner power flow for the decoupled AC networks between 12,000-bus system and 11,119-bus system.

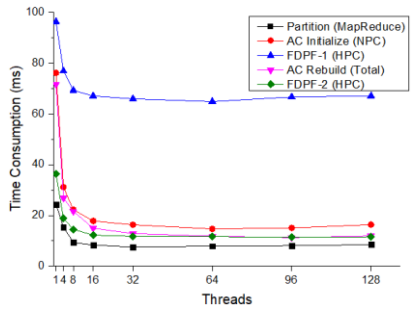

Figure 7(a). Time consumption for 12,000-Bus with multiple threads

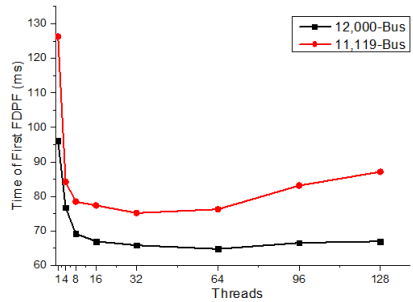

Figure 7(b). Comparison of FDPF time between 12,000-bus and 11,119-bus

As shown in Fig. 7(a), the performance of partition, NPC and HPC is improved from 1 to 32 threads. However, the time cost starts to increase and exceeds the gains of speed-up, when the number of running threads goes from 32 to 128. The limit factor for this result is that the bus size of the partitioned isolated area is 500, meaning that the parallel power flow cannot fully use the extra threads to do a complex calculation. To further reduce the computing time used in solving $Ax = B$, the LU matrices are stored after the first inner iteration since the topology of the admittance matrix maintains the same during calculation. Another factor that can affect the calculating performance is the size uniformity of each area. Regarding the Eq. (1), the sizes of partitioned areas are supposed to be close to reduce the waiting time between the largest and smallest areas. As shown in Fig. 7(b), the proposed method for the 11,119-bus system has better performance when fewer threads are in use, and the time increases quickly with more threads. The reason is that there are only 5 partitioned areas in the 11,119-bus system. Thus, the FDPF for each area could only be conducted with no more than 5 threads, and the smaller areas have to wait until the convergence of the largest area. The results of the artificial 12,000-bus system emphasized the improved efficiency with more areas with the same sizes. There is no waiting time in this test case, and more areas can be calculated at the same time with more threads.

V. CONCLUSION

The graph computation based sequential method is proposed in this paper to improve the efficiency of power flow analysis for a large-scale AC/DC system, with the employment of graph partition algorithm, NPC and HPC based parallel FDPF for multiple partitioned areas. The accuracy and efficiency are tested on the standard IEEE 300-bus system, the artificial 12,000-bus system, and the practical 11,119-bus system. The computation time is compared with Matlab using the same sequential method. It shows that for a hybrid system with over 10,000 buses, the proposed method can save about 75.5% computing time on average. In addition, further test results with multiple threads denote that the proposed method has good parallel performance. Besides, the detailed computing time of NPC and HPC also indicates that the number and sizes of partitioned areas have impacts on the efficiency.